\documentclass[reprint,amsmath,amssymb,aps]{revtex4-1}

\usepackage{graphicx}% Include figure files
\usepackage{dcolumn}% Align table columns on decimal point
\usepackage{bm}% bold math
\begin{document}

\preprint{APS/123-QED}

\title{A general solution to the preferential selection model}% Force line breaks with \\
% \thanks{To everyone and everything.}%

\author{
\firstname{Jake Ryland}
\surname{Williams}
}

\email{jw3477@drexel.edu}

\affiliation{
  Department of Information Science,
  College of Computing and Informatics,
  Drexel University,
  3675 Market St.
  Philadelphia, PA 19104}

\author{
\firstname{Diana}
\surname{Solano-Oropeza}
}

\email{ds3367@drexel.edu}

\affiliation{
  Department of Physics,
  Drexel University,
  32 S. 32$^\text{nd}$ St.
  Philadelphia, PA 19104}

\author{
\firstname{Jacob R.}
\surname{Hunsberger}
}

\email{jrh378@drexel.edu}

\affiliation{
  Department of Chemical and Biological Engineering,
  College of Engineering,
  Drexel University,
  3100 Market St.
  Philadelphia, PA 19104}

\date{\today}

\begin{abstract}
  We provide a general analytic solution to Herbert Simon's 1955 model for time-evolving novelty functions. This has far-reaching consequences: Simon's is a pre-cursor model for Barabasi's 1999 preferential attachment model for growing social networks, and our general abstraction of it more considers attachment to be a form of link selection. We show that any system which can be modeled as instances of types---i.e., occurrence data (frequencies)---can be generatively modeled (and simulated) from a distributional perspective with an exceptionally high-degree of accuracy.
\end{abstract}

\maketitle

%\tableofcontents
% describe the model as a general framework for selection processes

\section{\label{sec:intro} Preferential selection}
\vspace{-10pt}
What are the mechanistic processes through which social agents make selection decisions, or more concisely, how do people pick things? Social agents might express themselves through selection, and one well-known mechanism for understanding these processes comes from the study of complex networks---it is known as \textit{preferential attachment} \cite{barabasi1999a,krapivsky2001a}. It  traces its roots to the study of evolution~\cite{yule1924a}, and for text, its analog is well known as \emph{language generation}, i.e., an agent selects words~\cite{simon1955a}. This model for language came about in 1955 from the well-known social scientist Herbert Simon. It abstracts well to other selection contexts and can capture a variety of phenomena through modulation of its parameters~\cite{gerlach2013a,williams2015b,dodds2017a,williams2017a}, so we refer to this model more generally as \textit{preferential selection}.

We continue with the development and extension of this model through a generalized analysis for arbitrary, time-evolving \textit{novelty} rates, i.e, the capacity for the selection model to pick `new' things. Previously, general solutions were only available for this model when systems were assumed to obey a constant novelty rate. This has largely obstructed the applicability and usefulness of Simon's selection model to real-world data. Our solution provides a functional form for essentially all parameterizations, which will result in much greater applicability. Hence, future work will include experiments that seek to uncover insights from a fuller breadth of selection data from different social contexts. Likewise, computational tools for conducting these analysis will be fully developed for open-source release. \vspace{-10pt}

\subsection{Model setup}
\vspace{-10pt}
Preferential selection describes a sequence of $M$ instances (words): $(x_m)_{m=1}^M$ that have an onto relationship to a set of types $\mathcal{W}$ (a vocabulary). So, let $N = |\mathcal{W}|$ be the number of types and index the set of types (surface forms) $w_n\in\mathcal{W}$ by their order of appearance so that $w_n$ indicates the $n^\text{th}$ unique type in the stream. For any instance, $m$, define $n_m$ to be the number of types observed `so far', i.e., within: $\{x_k\}_{k=1}^{m}$. Let $\alpha$ (without subscript) denote a fixed novelty rate and let $\alpha_m$ denote one which varies by instance. Without loss of generality, when selecting instance $x_m$ Simon's model can be succinctly understood as a trade-off between two dynamics:
\begin{itemize}
    \item \textbf{exploitation}: with probability $\alpha_{m-1}$, the selected instance is a `new' (novel) type: $x_m = w_{n_{m-1} + 1}$
    \item \textbf{exploration}: with probability $(1 - \alpha_{m-1})$ $x_m$'s type is assigned randomly from $\{x_k\}_{k=1}^{m-1}$
\end{itemize}
Note: indexing requires $n_0 = 0$ (there are zero instances $\iff$ there are zero types) and $\alpha_0 = 1$ (the model has no false starts). \vspace{-10pt}

\subsection{Pre-existing solution}
\vspace{-10pt}
Now define $m_n: n=1,\cdots,N$ as the instance at which the $n\textsuperscript{th}$ type is introduced. In line with the rate-equations approach from \cite{dodds2017a}, consider the recursion equation: 
\begin{equation}
\label{eq:generalRate}
    f_m(w_n) = \left[1 + \frac{1 - \alpha_{m-1}}{m - 1}\right]f_{m-1}(w_n).
\end{equation}
Its expansion produces the following product:
\begin{equation}
\label{eq:generalProduct}
    f_m(w_n) = \prod_{j=m_n}^{m-1}\frac{j + 1 - \alpha_j}{j}.
\end{equation}
When $\alpha_m = \alpha$ for all $m$ (novelty is held constant), the numerator in the product produces $\Gamma$- and $\beta$-function representations:
\begin{equation}
    f_m(w_n) = \frac{\Gamma(m + \theta)\Gamma(m_n)}{\Gamma(m_n + \theta)\Gamma(m)} = \frac{B(m_n,\theta)}{B(m,\theta)}
\end{equation}
where $\theta = 1-\alpha$ denotes the exploitation probability (for convenience). In the latter, we substitute the Stirling approximation for $\beta$ functions, and arrive at an analytic frequency approximation:
\begin{equation}
    \label{eq:simonExact}
    \hat{f}(w_n) \approx\left(\frac{m_n}{m}\right)^{-\theta}.
\end{equation}

Following subsequent work~\cite{williams2017a}, for any $n\geq 1$ note that $\langle m_{n+1} - m_n\rangle$ is the expectation of a geometric distribution with success probability $\alpha$, i.e., that $\langle m_{n+1} - m_n\rangle = \alpha^{-1}$. This separates as: $\langle m_{n+1} - m_n\rangle = \langle m_{n+1}\rangle - \langle m_n\rangle$, and results in another recursion equation: $\langle n_{n+1}\rangle = \langle m_n \rangle + \alpha^{-1}$, which provides:
\begin{equation}
    \label{eq:Mestimate}
    \langle m_n \rangle
    = \frac{\alpha + n - 1}{\alpha}
    = \frac{n - \theta}{\alpha}
    .
\end{equation}

Approximation of $m_j$ by $\langle m_n\rangle = (n - \theta)/\alpha$ within the numerator and denominator of Eq.~\ref{eq:simonExact} renders the pre-existing form for the preferential selection model's analytic frequencies:
\begin{equation}
    \label{eq:simon}
    \hat{f}(w_n) \approx \left(\frac{n - \theta}{N - \theta}\right)^{-\theta}. \vspace{-10pt}
\end{equation}

\subsubsection{Non-constant novelty rates}
\vspace{-10pt}
Most works on the subject have acknowledged preferential selection supports non-constant novelty rates, but the topic has remained  largely unexplored in the literature. Some progress was made by~\cite{gerlach2013a}, though only focusing on specifically parameterized, power-law attenuating variation (with instance numbers). While this is empirically reasonable, the case was notably an example of a power-law-in/power-law-out phenomenon---assuming the power-law in the novelty function analytically produced a secondary power law in the resulting frequency distribution. However, general effects of non-constant novelty rates on frequency distributions are unknown (prior to our derivation, below).

Studies on the closely-related Growing Network (GN) model (and its variants) have focused on `attachment kernel' mechanisms \cite{krapivsky2001a}. But GN always selects, i.e., `links' a new type (node) to a pre-existing type (instance), for preferential selection this would be equivalent to a constant novelty rate of $\alpha = 0.5$, since a novel node is explored and an existing node is exploited with each link (instance). While it doesn't impact the resulting frequency distribution, the network picture has extra connectivity information that does not exist for preferential selection.

Our work (below) resolves two important limitations that have prevented the direct analysis of non-constant novelty rates. These are: 1) the challenges of adapting the rate-equations analysis to arbitrary novelty rates, and 2) a lack of data collection and/or representation of empirical novelty rates. We overcome both, and are thus able to formulate arbitrary frequency distributions implied by novelty rates through preferential selection. \vspace{-10pt}

\subsection{Solving for non-constant novelty}
\vspace{-10pt}
Starting from the novelty rate, $\alpha_m$, we have a quantity that generally varies with each observed instance.  But when \cite{gerlach2013a} explored evolving novelty rates, $\alpha_n$ was defined as function of $n$---the number of observed types, \emph{not} instances. Likewise, we will work with novelty rates that vary by types. But our modification of Eq.~\ref{eq:generalProduct}
assumes novelty varies as a step function that `steps' with new types, as opposed to each instance. In particular, we assume that the novelty rate can be written as $\alpha_m = \alpha_{n_m}$, for all $m$. Critically, if the $m+1^\text{st}$ instance is not a novel type, this assumption forces: $\alpha_m = \alpha_{m+1}$.

Substituting our step function into Eq.~\ref{eq:generalProduct} produces:
\begin{equation}
    f_m(w_n) = \prod_{k=n}^{n_m - 1}\prod_{j=m_k}^{m_{k+1}-1}\frac{j + 1 - \alpha_k}{j}.
\end{equation}
Since the $j$-indexed product has the same form as Eq.~\ref{eq:generalProduct}, i.e., $\alpha_k$ is constant with respect
to $j$, we can simply substitute the $j$-indexed product with a form analagous to the right hand side of Eq.~\ref{eq:simonExact}:
\begin{align}
    \label{eq:noveltyProduct}
    f_m(w_n) &= 
    \prod_{k=n}^{n_m - 1}\left(\frac{m_k}{m_{k+1}}\right)^{-(1 - \alpha_k)}\\
    &= \frac{m_n^{-(1 - \alpha_n)}}{m^{-(1 - \alpha_{n_m})}}\prod_{k=n}^{n_m - 2}\left(m_{k+1}\right)^{\alpha_{k+1} - \alpha_k}
\end{align}
and simplify (at right, above). \vspace{-10pt}

\subsubsection{This solution as a generalization}
\vspace{-10pt}
Provided $\alpha_k \rightarrow 1$ (novelty attenuates) smoothly, the above analysis implies the following frequency behavior:
\begin{equation}
    \label{eq:fLimit}
    f_m(w_n) \rightarrow \frac{m_n^{-(1 - \alpha_n)}}{m^{-(1 - \alpha_{n_m})}},\\
\end{equation}
which is quite similar to that of Eq.~\ref{eq:simonExact}. However, to move this to a frequency representation based on $n$ (the generalization of Zipf's law~\cite{zipf1935a}), we'll have to produce a separate approximation for $m_n$ based on the step-function novelty rate, $\alpha_n$, which we move onto next. But notably, we observe that
holding $\alpha_n$ to a constant, as $\alpha$, collapses Eq.~\ref{eq:noveltyProduct} into Eq.~\ref{eq:simonExact}, as any generalization should.

Focusing again on approximations, we consider how the inverse of the novelty rate---even when varying---still provides an estimate for the number of instances elapsed between observed novel types: $\alpha_n^{-1} \approx m_{n+1} - m_{n}$. This may be studied again through recursion:
\begin{equation}
    m_{n+1}\approx \alpha_n^{-1} + m_n = 1 + \sum_{k=1}^n\alpha_k^{-1}
    =1 + \sum_{k=1}^n\alpha_k^{-1}.
\end{equation}
Rearranging this equation, and noting the harmonic mean,  $\langle\alpha\rangle_n$, we are thus able to express:
\begin{equation}
    \label{eq:MGestimate}
    m_{n}\approx \frac{\langle\alpha\rangle_{n-1} + n - 1}{\langle\alpha\rangle_{n-1}}
\end{equation}
Note this form's resemblance to (and generalization of) the form presented in the middle of Eq.~\ref{eq:Mestimate}. This form relies only on the novelty rate (parameters, essentially) and the number of observed types ($n$), so we may utilize it by inserting for $m_{n}$ into Eq.~\ref{eq:fLimit} to obtain our approximation for the frequency scaling resulting from an arbitrary novelty rate:
\begin{equation}
    \label{eq:fScale}
    f_m(w_n) \propto \left(\frac{n + \langle\alpha\rangle_{n-1} - 1}{\langle\alpha\rangle_{n-1}}\right)^{-(1 - \alpha_n)}
\end{equation}
(under the condition that novelty attenuates smoothly). But Eq.~\ref{eq:fScale} is not the most desirable form for empirical exploration, and rather best used to assess familiar functional characteristics. For a more accurate functional form that doesn't depend on smooth novelty attenuation, one need only substitute Eq.~\ref{eq:MGestimate}'s approximation into Eq.~\ref{eq:noveltyProduct}, which is exact and amenable to computation and data, as our investigation continues into.

To close this section and highlight one last aspect of generalization, we consider the specific, well-known novelty function from \cite{gerlach2013a}. Letting $\alpha_n$ be a function of types with a power-law attenuation occurring after a break point, $b\geq1$, we assume the rate of novelty's attenuation is controlled (after $b$) by a negative scaling, $\mu\geq 0$: 
\begin{equation}
    \label{eq:amix}
    \hat\alpha_n (\alpha_0, \mu, b) =
    \begin{cases} 
        \alpha_0                               &  n\leq b \\
        \alpha_0 \left(\frac{n}{b}\right)^{-\mu}  &  n > b 
    \end{cases} 
\end{equation}
Where $n\leq b$ the novelty function remains a fixed constant, $\alpha_0$, and afterwards it attenuates as a power law, in and of itself. Inserting Eq.~\ref{eq:amix} into Eq.~\ref{eq:fScale}, the Euler-Maclaurin formula provides $\langle\hat\alpha\rangle_n\rightarrow \frac{\alpha_n}{1+\mu}$ which allows us to note the limiting proportionality:
$$\lim_{n\rightarrow\infty}f_m(w_n) \propto (n-1)^{-(1+\mu)},$$
which is equivalent to the same form derived differently in other work~\cite{gerlach2013a}. \vspace{-10pt}

\section{Empirical novelty rates}
\vspace{-10pt}
Here, we derive and explore several different possible methods for producing an empirical, $N$-parameter novelty rate from data, $(\hat{\alpha}_n)_{n=1}^N$. \vspace{-10pt}

\subsection{Reading order}
\begin{figure}[t!]
  \includegraphics[width=0.5\textwidth, angle=0]{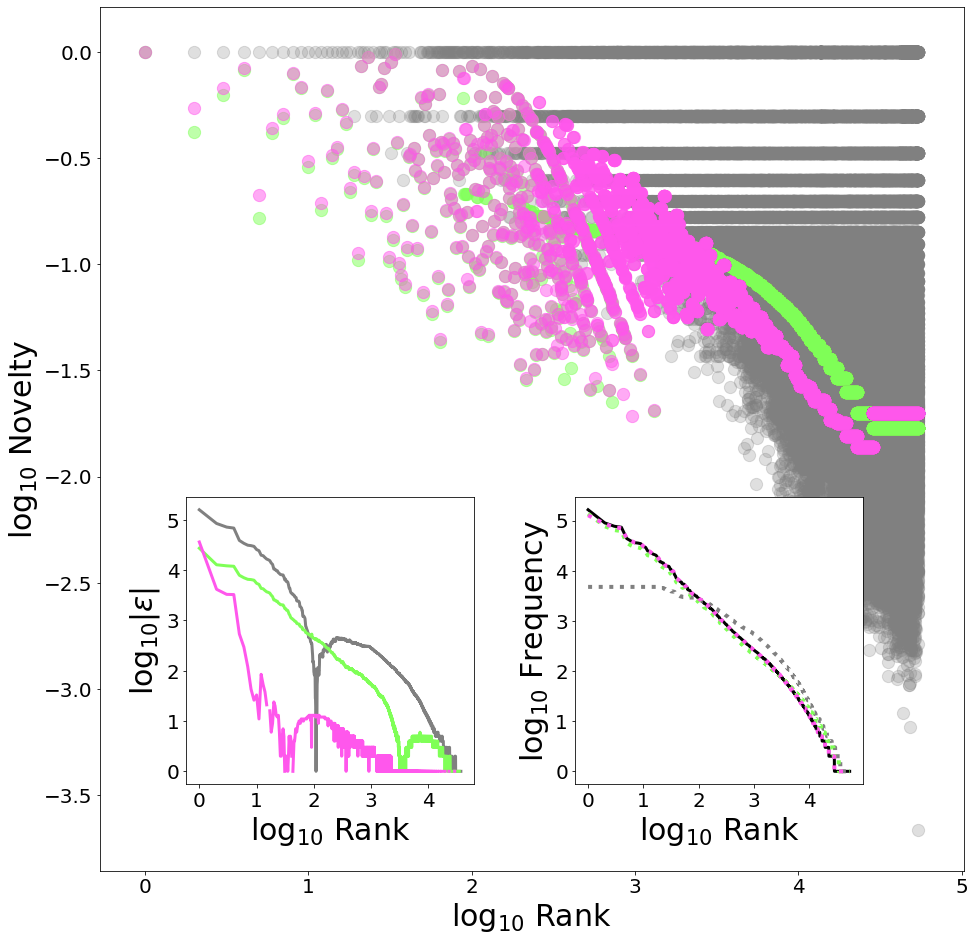}
  \caption{
        Empirical novelty rates are presented for the collected works of Georg Ebers.
  }
  \label{fig:EmpiricalNoveltyFits}
\end{figure}

\vspace{-10pt}
Inspired by the language context, this \emph{cognitively na\"ive approach} focuses on computing the gap sizes, $\Delta_n$ (in number of selection instances), observed between the selection of novel types. Their reciprocals produce a noisy representation of the novelty function: $\hat{\alpha_n} = 1/\Delta_n$. A diagram is presented as Eq.~\ref{eq:empiricalAlpha} to illustrate this:
\begin{equation}
\label{eq:empiricalAlpha}
\underbrace{A\:}_{\Delta_1}
\underbrace{B\:A\:}_{\Delta_2}
\underbrace{C\:B\:A\:B\:}_{\Delta_3}
\underbrace{D\cdots}_{\Delta_4}
\end{equation}
This empirical novelty rate offers contextualized information about a selection's presented order, but without connection to frequency accumulation. Its noise is an exhibition of context, i.e., the subtly non-random order of empirical selection, or in a more modern parlance for language~\cite{vaswani2017a}, the attention of word choice to context. In Fig.~\ref{fig:EmpiricalNoveltyFits}, an example of the reading-order representation (and some immediate limitations to its value) for a large document can be seen. Like other documents, reading-order novelty (grey points, main axis) can be seen to transform through Eq.~\ref{eq:noveltyProduct} into a rank-frequency distribution (grey dotted, lower right inset) that is extremely out of sync with the empirical frequencies, except perhaps at low frequencies. \vspace{-10pt}

\subsubsection{Interpreting the reading order novelty}
\vspace{-10pt}
If non-constant novelty rates can produce accurate rank-frequency distributions for language, they will have to be constructed out of sync with the way an author presents a document. We do so in the next sections, but this finding in and of itself is, empirically, critical. If the reading order doesn't capture the nature of accumulation, then what is the model failing to capture? It is possible that modeling an additional \textit{scrambling} (shuffling) process could overcome this limitation. This would regard reading order as some kind of a useful randomization of selected instances---perhaps an artifact of the regularizations of language we experience through syntax, grammar, and semantics. But ultimately, more work is required to extract value from this empirical representation. \vspace{-10pt}

\subsection{Birthday-derived novelty representations}
\vspace{-10pt}
Both of the following representations fit tightly to frequency because their formulation is directly based on the empirical frequencies. Both, however, descend from a central assumption about \textbf{birthdays}. These instance numbers, $m_n$, describe the total progress the system makes before the $n^\text{th}$ type emerges. Critically, \emph{assuming} types appear in rank-frequency order, the proportional selection property assures that 
\begin{equation}
    \label{eq:mnemp}
    m_n\approx \left.\frac{m}{f(w_r)}\right|_{n=r}
\end{equation}
This can be understood as follows: between birthdays the model guarantees proportional selection events---\emph{only}. At the $n^\text{th}$ birthday, the $n^\text{th}$ type's frequency is 1, so it is guaranteed that the relative proportion with any other (lower-ranked) type, i.e., with $k < n$ will hold:
\begin{equation}
    \frac{f_{m_n}(w_k)}{f_{m_n}(w_n)} \approx \frac{f_{m}(w_k)}{f_m(w_n)}
\end{equation}
So this, and the fact that $f_{m_n}(w_n) = 1$ ensure the result:
\begin{equation}
    m_n = \sum_{k=1}^n{f_{m_n}(w_k)} \approx \sum_{k=1}^n{\frac{f_{m}(w_k)}{f_m(w_n)}}. \vspace{-10pt}
\end{equation}

\subsubsection{Boundary pivoted}
\vspace{-10pt}
As we know from \cite{dodds2017a}, the first-appearing type introduced to a preferential selection system holds a special, distributional position. Proportionally, it is the most distributionally stable type (with respect to frequency and analytic limits) that the model produces, so it forms a useful pivot. When the novelty rate is constant (Eq.~\ref{eq:simonExact}), one can apply basic logarithmic equation solving:
\begin{equation}
    \hat{\alpha}\approx 1 - \frac{\log(f(w_1)/f(w_N))}{\log(m_1/m)}
\end{equation}
where we have used the fact that $f(w_N)=m_1=1$ to highlight that this approximation characterizes novelty for the range of types (all of them) that were `born' under this (constant) rate. 

As a time-evolving novelty-rate increases the parameters (degrees of freedom), solving for these requires more information---all of the frequency distribution. But because selection is proportional, we can leverage this same formulation over the steps of our $\alpha_n$ step function. Using the fixed proportions of types born immediately adjacent to one another, we can leverage a cancellation of factors in Eq.~\ref{eq:noveltyProduct} to derive:
\begin{equation}
    \hat{\alpha}_n\approx 1 + \frac{\log(f(w_{n-1})/f(w_{n}))}{\log(m_{n-1}/m_n)},
\end{equation}
This now characterizes the novelty step for the ($n^\text{th}$) range of types that were `born' under this rate. An example novelty function derived from this analytic cancellation is likewise presented in Fig.~\ref{fig:EmpiricalNoveltyFits} as the fit in green, which at this point produces a frequency prediction (bottom right) that is strong enough to be challenging to distinguish from the empirical frequencies by eye. In the lower left, we see it exhibits roughly an order of magnitude less error per word than reading-order novelty (a vast improvement). \vspace{-10pt}

\subsubsection{Proportioned gaps}
\vspace{-10pt}
This representation also utilizes the `birthday' approximations for $m_n$, but does so in a more-ad hoc fashion. Using only Eq. \ref{eq:mnemp}, we simply note that by definition $\Delta_n = m_{n} - m_{n - 1}$ with 
\begin{equation}
    \hat{\alpha}_n \approx (\Delta_n)^{-1} = (m_{n} - m_{n - 1})^{-1}.
\end{equation}
An example novelty function derived from proportioned gaps is likewise presented in Fig.~\ref{fig:EmpiricalNoveltyFits}. There, it is the `best fit' (in pink), exhibiting roughly an order of magnitude less error than the boundary-pivoted form. \vspace{-10pt}

\subsubsection{Interpreting birthday-derived novelty}
\vspace{-10pt}
Both birthday-derived formulations have a starting condition, i.e., are only defined for $n > 1$. But since the first type is guaranteed to appear as the first instance ($\alpha_0 = 1$), $\alpha_1$ only meaningfully defines well the novelty rate \emph{after} the first type's appearance. So the model \textit{and} these representations are all forward-looking---Eq.~\ref{eq:noveltyProduct} technically doesn't utilize any `last' value $\alpha_N$ as a result of the indexing. Additionally, both approximations are subject to finite size effects as a result of integer frequencies in data.  Particularly, there are many low-frequency types that have large, ambiguous ranks. So both representations are best computed by batching same-frequency types in aggregate calculations, i.e., treating each plateaux in the rank-frequency distribution as though its types appeared under a single `step' of constant novelty. This aligns to these types' indistinguishably by frequency. \vspace{-10pt}

\begin{figure}[t!]
  \includegraphics[width=0.5\textwidth, angle=0]{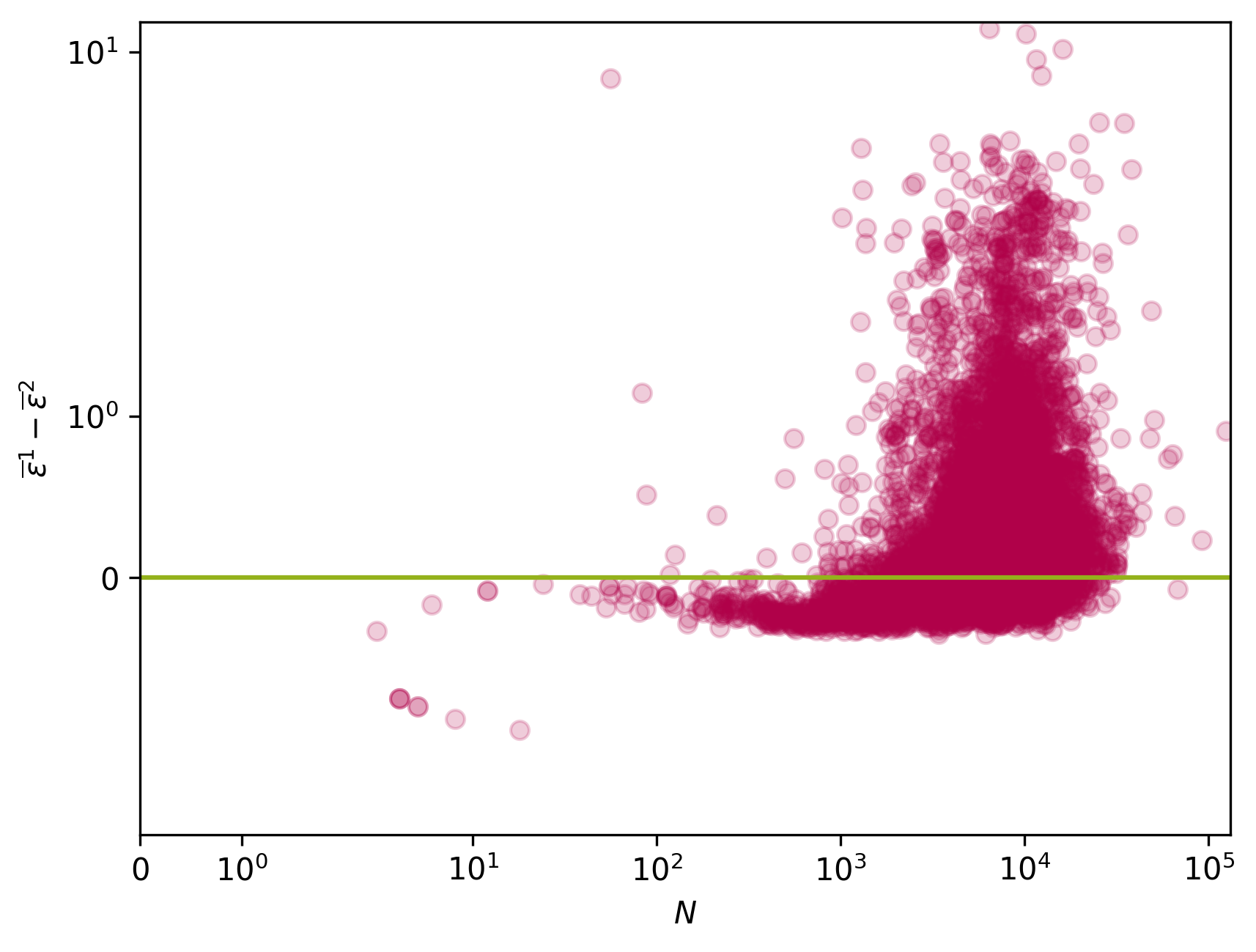}
  \caption{
        Comparison of the performance of birthday-derived novelty representations as a function of vocabulary size, $N$.
  }
  \label{fig:NoveltyComparisons}
\end{figure}

\subsection{Evaluating novelty representations}
\vspace{-10pt}
Considering how wildly divergent the reading order representation is from corresponding empirical frequencies, this exploration compares the two birthday-derived novelty representations. Hence, we present a preliminary empirical characterization of the circumstances in which these representations form better and worse models---it turns out, neither is generally, empirically `best'. In particular, across the Project Gutenberg eBooks collection (over $20,000$ documents, from a variety of languages and topical sources) we: 1) compute both birthday-derived representations, $\hat{\alpha}^1, \hat{\alpha}^2$; 2) apply them through Eq.~\ref{eq:noveltyProduct} to produce corresponding frequency representations, $\hat{f}^1, \hat{f}^2$; and 3) compute the average of absolute error for each, $\overline{\varepsilon}^1, \overline{\varepsilon}^2$; where the superscripts indicate the analytic boundary-pivoted (1) and empirically-proportioned gaps (2) representations. 

While both birthday-derived novelty representations produce models that fit tightly to frequency (see Fig.~\ref{fig:EmpiricalNoveltyFits} insets), we note that neither (according to derivation) appears to be objectively better (across the eBooks). The result of this experiment's application can be see in Fig. \ref{fig:NoveltyComparisons}, which presents $\overline{\varepsilon}^1 - \overline{\varepsilon}^2$ as a function of vocabulary size, $N$. Interestingly, the proportioned gaps commonly outperform the boundary-pivoted representation (most documents fall below the green line). But for larger documents (more than $10^3$ terms), a second regime emerges, where the boundary-pivoted representation begins to outperform the proportioned gaps. In particular, for just some large documents (never small), the proportioned gaps fail dramatically. This interestingly identifies two clusters of documents, with one being characterized by some large-scale effect. Further investigation characterizing this variation is thus warranted.\vspace{-10pt}

\section{Future work}
\vspace{-10pt}
This work is the beginning of an investigation that is now being directed towards empirical work. With the analytic solution in place and a number of accurate, empirical novelty representations, we will move on to regressing low-parameter characterizations of non-constant novelty rates, such as Eq.~\ref{eq:amix}. As we've studied in other work, regressing these parameters ($\mu$) can produce a potent featurization for understanding qualitative characteristics of language generators, e.g., social bots~\cite{clark2016a}. Being able to regress directly from well-representing novelty distributions will provide a large boost to performance at detection. The generality of this solution and ability for modeling arbitrary novelty-evolving context likewise allows for this model's applicability to a diversity of categorical data streams. Hence, we intend to investigate the capacity for these modeling approaches to effectively describe the selection characteristics of other social selection processes that we expect to be strongly modulated by proportional attention towards historical frequency, such as Twitter users liking tweets, authors citing papers, or journalists referencing social media users. \vspace{-10pt}

\section*{Acknowledgments}
\vspace{-10pt}
This document is based upon work supported by the National Science Foundation under grant no. \#1850014.

\bibliography{main}% Produces the bibliography via BibTeX.

\end{document}